\documentclass[12pt]{article}
\usepackage[usenames,dvipsnames]{color}
\usepackage{graphicx}
\usepackage{amssymb}
\usepackage{breqn}
\usepackage[font=small,labelfont=bf]{caption}

\title{Kinetic description of wave induced plasma flow in the radio frequency domain}
\author{Dirk Van Eester, K. Cromb\'{e}, Ye.O. Kazakov \\
Laboratorium voor Plasmafysica - Laboratoire de Physique des Plasmas \\
Association ``EURATOM - Belgian State'' \\
Trilateral Euregio Cluster\\
Renaissancelaan 30 Avenue de la Renaissance\\
B-1000, Brussels, Belgium \\
\vspace{-1cm}
\date{November 29, 2013}
}
\begin{document}

\maketitle

\abstract{
A model for ICRH induced flows in the presence of a strong magnetic field is presented. These flows are the finite temperature counterpart of flows existing in cold plasmas described e.g. in [D. Van Eester et al., \textit{Plasma Phys. Control. Fusion} \textbf{55} (2013) 025002] and thus do not rely on the waves being damped. The kinetic corrections offer insight in what happens at cyclotron resonances. Authors commonly either rely on the confining magnetic field $\vec{B}_o$-field to be strong, or the electric field $\vec{E}$-field to be rapidly varying but are not accounting for both when writing down the solution of the equation of motion on the slow time scale. In this paper, the equation of motion is solved for constant $B_o$ to keep the discussion as simple as possible. The simultaneous presence of $\vec{B}_o$ and the $\vec{E}$-field inhomogeneity causes drifts perpendicular to the $\vec{B}_o$ and to other slow time scale accelerations, the Ponderomotive acceleration being one of them. Because of the first and having tokamak applications in mind, these flows - although small in magnitude - cause drifts that enter in competition with transport induced flows.
}

\section{Introduction}

Flow induced by ion cyclotron resonance frequency (ICRH) or radio frequency (RF) waves has been observed in various tokamaks \cite{Chan, Eriksson, Zhang, DVE_He3}. In particular, mode conversion heating scenarios seem promising to induce flow drive since the localization where the flow velocity is maximal coincides crudely with the mode conversion layer(s), suggesting that - between the point where it is excited and the (usually nearby) point where it is absorbed - the short wavelength ion Bernstein or ion cyclotron branch is instrumental to drive flow efficiently. Although rigorous modeling of this effect in principle requires accounting for the fact that the first moment of the distribution function is finite (an assumption that is at odds with the assumption commonly made in wave models that the relevant distributions are non-shifted Maxwellians i.e. that the populations are close to being in thermal equilibrium and do not exhibit net drifts), the observed (poloidal) velocities are in the range of a few $km/s$ and thus are reasonably small w.r.t. the thermal velocities of charged particles in the core of hot tokamak plasmas. Hence, a first assessment can be made developing a theory that discusses RF induced momentum transfer assuming the distribution is a simple Maxwellian and omitting the back-reaction on the distribution.

Selfconsistently accounting for the interaction between electromagnetic fields and charged particles not only requires solving the wave equation (yielding the field pattern for a given set of distribution functions of the various plasma constituents) and the Fokker-Planck equations (yielding the distribution functions for a given wave pattern) but equally forces one to trace the modifications on the even slower (transport) time scales in the problem. Since the various species Coulomb collisionally interact with each other, energy and momentum is transferred between them. The simultaneous action of the waves and the Coulomb interaction deforms the distributions and makes them non-Maxwellian. From the first velocity moments of the distribution the populations' drifts and thus the currents can be deduced. These velocities need to be fed into the continuity equation to assess the RF induced density depletion, and the thus obtained densities need in turn to be accounted for in the slow time scale Maxwell equations or - in the easiest possible case - the Poisson equation to find out which electric field that tries to restore charge neutrality is auto-generated by the plasma. Solving the problem as a whole in the volume of large machines such as JET is a gigantic task which is far from being realizable trustworthily at present. In the present paper only 1 subaspect of this problem is addressed: that of the determination of the wave induced net drifts.

The bottleneck of the computation is that various aspects of the physics occur on very different time scales, which all need to be accounted for: The wave-particle interactions happen on the time scale imposed by the antenna frequency, while the fate of the energy (and momentum) passed on to the particles requires accounting for the orbits of the particles (and thus brings in the bounce time scale) as well as the cross-talk between particle populations (which occurs on diffusion time scale). The Fokker-Planck equation accounts for the slow time scale physics via the quasilinear diffusion operator which uses the result of the linear response and feeds it back into the evolution equation for the distribution function which gives rise to a second order correction which has a rapidly varying component as well as a component that only varies on the slower time scales. Performing the averages over \textit{all} fast time scales this yields the quasilinear correction to the slow time scale evolution equation. In the case of a charged particle in a tokamak there are 4 fast aspects of the motion. They correspond to the time scale imposed by the antenna driver frequency, and those of the cyclotron, the bounce and the toroidal drift motions (see e.g. \cite{Kaufman}; there is a vast amount of literature on this subject). Transport causes macroscopic quantities to be modified, and the here discussed term is one of the features affecting dynamics on the slowest time scale.

The here presented approach is based on the philosophy adopted to derive the Ponderomotive force. Not surprisingly, this way of looking at the problem is - when done fully rigorously - intimately related to the approach discussed in the previous paragraph: It seeks the solution for the linear response and substitutes the result into the second order correction term; averaging over the fast time scales then equally yields a correction term to the slow time scale dynamics.

This short text does not aim at a full description of ICRH induced rotation but limits itself to the study of the generalization to hot plasmas of the cold plasma Ponderomotive force term believed to be responsible for ICRH induced density modifications in front of the antenna \cite{DVE_Ne_depletion}.

\section{Derivation of the relevant equation}

\subsection{Cold plasma expression for the ICRH induced drift}

The equation of motion of a set of charged particles in a plasma is
$$m\frac{d}{dt}\vec{v}= q \vec{E} e^{-i \omega t} +  q \vec{v} \times \vec{B}_{o} + \vec{F}_{slow}$$
where $m$ is the mass, $T$ the temperature, $q$ the charge and $\vec{v}$ the velocity of the flow of one of the types of particles in the plasma. $\vec{F}_{slow}$ represents the forces other than the explicitly retained Lorentz forces due to the strong but static magnetic field $\vec{B}_o$ and the relatively small but rapidly varying electric field, driven at the externally imposed frequency $\omega$. In its simplest form $\vec{F}_{slow}=-N^{-1}\nabla NkT$, where $N$ is the density and $T$ the temperature, but a more complete description includes terms such as the anisotropic pressure corrections. These forces are assumed to be acting on a time scale that is long compared to that of the motion forced by the confining magnetic field and the driven wave source. The above can more compactly be rewritten as
$$\frac{d}{dt}\vec{v}= \vec{\epsilon} e^{-i \omega t} + \Omega \vec{v} \times \vec{e}_{//} + \vec{a}_{slow}.$$
Here $\vec{\epsilon}=q\vec{E}/m$, $\Omega=qB_o/m$, $\vec{e}_{//}=\vec{B}_{o}/B_o$ and $\vec{a}_{slow}=\vec{F}_{slow}/m$. Note that the $driven$ magnetic field has been neglected since it usually is a small correction to the terms retained. Its effect will be added later in the text. We will further adopt a quasi-homogeneous approach and take the static magnetic field $\vec{B}_o$ to be straight and homogeneous, hereby implicitly assuming that the scale length on which the waves vary is much shorter than that of the static magnetic field. This is a major simplification but it allows to limit the algebra to a bare minimum at the price of missing some of the effects brought about by the non-homogeneity of the considered equilibrium. Anticipating that the motion can be split into a fast dynamics part that responds on the time scales imposed both by the generator and the strong confining magnetic field, and a slow dynamics part which accounts for remaining net drifts when averaging over all fast time scales, we split both velocity $\vec{v}=\vec{v}_o+\vec{v}_1$ and position $\vec{x}=\vec{x}_o+\vec{x}_1$ into a zero order and a perturbed part. The electric field is inhomogeneous in general. Assuming the perturbation is small and thus following the usual philosophy for the derivation of the Ponderomotive force, $\vec{\epsilon}$ can be approximated by the truncated Taylor series expansion
$$\vec{\epsilon}(\vec{x})\approx\vec{\epsilon}(\vec{x}_o)+(\vec{x}-\vec{x}_o).\nabla \vec{\epsilon}(\vec{x}_o)$$
so that the fast dynamics is governed by the equation
$$\frac{d}{dt}\vec{v}_1=  \vec{\epsilon} (\vec{x}_o)e^{-i \omega t} + \Omega \vec{v}_1 \times \vec{e}_{//},$$
while the slow dynamics are captured by
$$\frac{d}{dt}\vec{v}_o=\vec{a}_{Pond} + \Omega \vec{v}_o \times \vec{e}_{//} +\vec{a}_{slow}=\frac{1}{2}Re[ <(\vec{x}^*-\vec{x}_o^*).\nabla \vec{\epsilon}(\vec{x}_o) e^{-i \omega t}>] + \Omega \vec{v}_o \times \vec{e}_{//} +\vec{a}_{slow}$$
with $<...>$ representing the averaging over all fast oscillations. Solving the equation of motion for the fast dynamics starting at a reference time $t_o$ yields
$$x_{\perp_{1}}=x_{\perp_{1},o}+\frac{1}{2} \Big[  \frac{v_{+o}}{-i\Omega} e^{-i \Omega \tau} + \frac{v_{-o}}{i\Omega} e^{i \Omega \tau} \Big ] - \frac{\omega \epsilon_{\perp_{1}} + i \Omega \epsilon_{\perp_{2}}}{\omega (\omega^2-\Omega^2)} e^{-i\omega \tau},$$
$$x_{\perp_{2}}=x_{\perp_{2},o}+\frac{1}{2i} \Big[  \frac{v_{+o}}{-i\Omega} e^{-i \Omega \tau} - \frac{v_{-o}}{i\Omega} e^{i \Omega \tau} \Big ] - \frac{\omega \epsilon_{\perp_{2}} - i \Omega \epsilon_{\perp_{1}}}{\omega (\omega^2-\Omega^2)} e^{-i\omega \tau},$$
$$x_{//}=x_{// o} + v_{// o} \tau -\frac{\epsilon_{//}}{\omega^2}e^{-i\omega \tau}.$$
where $\tau=t-t_o$, the factor $exp[-i\omega t_o]$ has been absorbed in $\vec{\epsilon}$ for abbreviating the notation and

\textcolor{black}{$$v_{//o}=v_{//}(\tau=o)-\frac{i\epsilon_{//}}{\omega}.$$}

The motion perpendicular to the static magnetic field contains a constant contribution, an oscillatory contribution that responds to the cyclotron motion imposed by the static magnetic field, and a driven oscillatory contribution that corresponds to the movement caused by the electric field. Since the magnetic field does not exert any influence parallel to its field lines, the parallel motion is only constituted of the guiding center drift along the magnetic field line and resulting from the initial parallel velocity $v_{// }$, and the driven response to the electric field.

In tokamaks, the magnetic field is of the order of a few Tesla, while the electric field excited by RF antennas is commonly of the order of a few $10^4V/m$ near the antenna and decaying as it moves towards the plasma core. Taking the thermal velocity as a measure for the typical velocities reached, the velocity is of order $10^6m/s$ for typical temperatures of order $10keV$. Hence, in the hot plasma core $|E/[vB_o]| << 1$ holds.

With the obtained expressions, the Ponderomotive term in presence of a strong magnetic field can now be evaluated. Only retaining net non-oscillatory terms one finds \cite{DVE_Ne_depletion}
\begin{equation} \label{MAIN_EQ}
\vec{a}_{Pond}=-\frac{1}{2}Re \Big [ +\frac{1}{\omega(\omega^2-\Omega^2)} \Big ( [\omega \epsilon_{\perp_{1}}^* - i \Omega \epsilon_{\perp_{2}}^*]\frac{\partial}{\partial x_{\perp_{1}}} \vec{\epsilon} + [\omega \epsilon_{\perp_{2}}^* + i \Omega \epsilon_{\perp_{1}}^*]\frac{\partial}{\partial x_{\perp_{2}}} \vec{\epsilon}\Big ) $$
$$ + \frac{1}{\omega^2} \epsilon_{//}^* \frac{\partial}{\partial x_{//}} \vec{\epsilon}
-   \frac{i v_{//o}^*}{\omega} \frac{\partial}{\partial x_{//}} \vec{\epsilon}   \Big],
\end{equation}
in which the functions appearing in the cold plasma dielectric tensor elements (see e.g. \cite{Swanson,Stix}) can be recognized and in which the last term only appears in case there is a finite zero order parallel drift velocity $ v_{//o}$ in absence of the RF field. Note that this  term is rapidly varying since $\vec{\epsilon}\propto exp[-i\omega t_o]$, while the other terms are time independent. As the Ponderomotive effect is a slow time scale rather than a fast time scale term and interpreted as corrective force felt along the orbit, it can be neglected by either considering a 'running' average ($\vec{x}-\vec{x}_o \rightarrow \vec{x}-\vec{x}_o - v_{//,0}\tau \vec{e}_{//} $) and/or performing a supplementary average on the fast time scale to take an average on all possible initial phases when performing the time average. Only considering the direct response to the driver frequency $\omega$ when solving the fast equation of motion (i.e. solving the equation of motion on the fast time scale by directly assuming $\vec{v}_1\rightarrow i\omega \vec{v}_1$), Klima \cite{Klima} derived a similar expression.

Putting the obtained expression for the Ponderomotive acceleration into the equation of motion for the slow dynamics, the latter can also be solved in a similar way as the equation of the fast dynamics. Note that - although none of the $forces$ varies on a rapid time scale - this equation of motion $still$ contains very fast variations, notably those imposed by the confining magnetic field. Hence time averaging over a cyclotron period will still be required to isolate the actual slow dynamics. First dropping the acceleration $\tilde{\vec{a}}$ due to the dynamics other than those imposed by the confining field allows to find the cyclotron oscillation solution
$$v_{\perp}=v_{\perp,o}=constant$$
$$\phi=\phi_o-\Omega \tau.$$
Subsequently adopting the method of the variation of the constants yields the equations
$$\frac{dv_{\perp,o}}{dt}=cos \phi \tilde{a}_1 +sin \phi \tilde{a}_2$$
$$\frac{d\phi_{o}}{dt}=-v_{\perp,o}\Omega -sin \phi \tilde{a}_1 +cos\phi \tilde{a}_2.$$
Since the cyclotron gyration dominates the slow motion, this set can be solved to leading order by using the cyclotron oscillation solution in the right hand side of these equations. This adds the slow time scale contribution
$v_{o,\perp 1}=+\tilde{a}_{\perp 2}/\Omega$, $v_{o,\perp 2}=-\tilde{a}_{\perp 1}/\Omega,$
to the solution. This contribution is of the usual form
$$\vec{v}_{drift,\perp}=\frac{\vec{a}\times \vec{e}_{//}}{\Omega}=\frac{\vec{F}\times \vec{B}_o}{q B_o^2}$$
of drift velocities in presence of a strong magnetic field together with another force $\vec{F}=m\vec{a}$. As the magnetic field does not influence the parallel dynamics, this drift velocity is $perpendicular$ to the confining field while the parallel dynamics is identical to the case in absence of $\vec{B}_o$, supplemented by a correction due to the parallel electric field.

Note that the slow time scale drift velocity obtained here is completely different in nature from the velocity obtained when $neglecting$ the effect of the strong magnetic field when integrating the slow time scale equation of motion. For example, Myra \cite{Myra2002} obtained a steady state solution of the slow time scale equation of motion by balancing the slow time scale (Ponderomotive + other relevant) accelerations with a friction term i.e. by solving
$$\frac{d\vec{v}_o}{dt}=\vec{a}_{slow}-\nu_{friction} \vec{v}_o$$
imposing $d/dt=0$. One readily finds
$$\vec{v}_o=\frac{\vec{a}_{slow}}{\nu_{friction}}.$$
Aside from the fact that the obtained drift velocity is $aligned$ with the acceleration whereas the drift in presence of a strong field is perpendicular to it, this result differs from the one presented in the present paper in the magnitude of the obtained velocity: Since the characteristic time scale on which friction takes place is typically much longer than the cyclotron period, the obtained drift velocities - for a given $\vec{a}_{slow}$ - are much larger. If $\nu_{friction}$ is interpreted as a collision frequency, the 2 obtained velocities are typically 4 orders of magnitude apart. In case it is to be interpreted to be connected to the Coulomb collisional slowing down time, the results are another 4 extra orders of magnitude apart. Which of the 2 expressions is the more relevant one? The present paper studies wave-particle interaction in a strong but homogeneous magnetic field. In that case the equations of motion - both on the fast and slow time scales - can readily be solved analytically provided the electric field is sufficiently small so that it makes sense to use the orbit unperturbed by the RF fields as a first, reasonably good approximation to find an improved expression for the gyro phase and the magnitude of the perpendicular velocity. This in turn allows to isolate 'fast' from 'slow' variations in known forms ('fast' Larmor oscillation and 'slow' drifts perpendicular to a 'slow' force and the confining magnetic field). In the literature various forms of 'slow time scale' equations can be found. When deriving transport equations in presence of a strong magnetic field, it is customary to average over the fast cyclotron gyration and find expressions for how quantities vary along guiding center orbits (see e.g. \cite{BalescuBooks}). In the thus obtained equation of motion of the guiding center, the $q \vec{v}_o \times \vec{B}_o$ force term giving rise to rapid oscillation around the $\vec{B}_o$ magnetic field lines is absent and the guiding center velocity is the sum of a number of terms of the earlier found form $\vec{v}_{drift}=\vec{a}\times \vec{e}_{//}/\Omega$. When focussing on fastly driven electric fields (see e.g. \cite{Klima}), terms that oscillate at the driver frequency no longer appear in the 'slow time scale' equation but the $q \vec{v}_o \times \vec{B}_o$ force term $is$ present and has to be dealt with. Clearly, for a sufficiently strong $\vec{B}_o$ field this will result in fast variations which will affect the - supposedly - slow motion in a dominant way. To formulate the proper slow time scale equation of motion both in presence of a strong and inhomogeneous $\vec{B}_o$ and driven, rapidly varying, inhomogeneous electric fields, a combination of these 2 philosophies is required. Such an approach is beyond the scope of this paper. It is the authors' opinion, however, that it is unlikely that the effect of the strong magnetic field will be dominated by a friction effect many orders of magnitude weaker, and that the here derived result for rapidly varying electric fields combined with a strong but homogeneous field is at least indicative of phenomena occurring in a more generalized description. For that reason, it will be assumed that the modest $\vec{B}_o$ induced drifts of order $|\vec{a}_{Pond}/\Omega|$ are physically relevant while the demonstration that much larger drifts of order $|\vec{a}/\nu|$ can (or cannot) be induced is left for later  scrutiny when a more complete model for the slow time scale dynamics is at hand.

Although small with respect to the thermal velocity, the retained drifts can play a non-negligible role if they induce net drifts in directions that are out of reach in absence of RF perturbations. The obvious candidates for potentially important drifts that come to mind when thinking in terms of the here adopted simple model where guiding centers are glued to magnetic field lines in absence of perturbing electric fields are drifts $across$ magnetic surfaces. In the next section, a hot plasma generalization of the results so far obtained is derived.

\subsection{Hot plasma generalisation}

The obtained expressions diverge when the cyclotron frequency approaches the driver frequency. The obtained expressions can readily be upgraded by accounting for the fact that not only $exp[-i\omega t]$ but also the electric field amplitude itself varies with time when accounting for the spatial dependence of $\vec{\epsilon}$ on the time via the time dependent position. Adopting the usual approach of expressing quantities in terms of the guiding center to isolate the gyromotion from the other motions and assuming that the field locally varies as $\vec{E}=\vec{E}_o exp[i \vec{k}.\vec{x}]$ (as opposed to $\vec{E}=\int d\vec{k} \vec{E}_{\vec{k}}exp[i\vec{k}.\vec{x}]$ containing contributions of the full spectrum $-\infty \le k_{\beta} \le +\infty$), the expression for the electric field is now
$$\vec{\epsilon}=\frac{q}{m}\vec{E}=\sum_N J_N(\xi)exp[-i\tilde{\omega}_N \tau] \vec{\epsilon}_{o,N}$$
in which $\tilde{\omega}_N=\omega-N\Omega-k_{//}v_{//}$, $\xi=k_\perp v_\perp/\Omega$ is the argument of the Bessel functions $J_N$, $\tau=t-t_o$ and $\vec{\epsilon}_{o,N}=\vec{\epsilon} exp[-i(\omega t_o+N\phi_o-N\Psi)]$; $N$ is the cyclotron harmonic. Note that all time dependence has been located in the phase $exp[-i\tilde{\omega} \tau]$. This is achieved in the usual way (see e.g. \cite{Stix}) by writing the oscillatory phase of the field as a sum of contributions highlighting the weight of each of the cyclotron harmonics. The gyrophase $\phi$ varies as $\phi=\phi_o-\Omega\tau$ and $\vec{k}=k_\perp [\vec{e}_{\perp_{1}}cos\Psi+\vec{e}_{\perp_{2}}sin\Psi]+\vec{e}_{//}k_{//}$. In the present - simplified - description the guiding center motion is simply the uniform motion along the magnetic field lines at constant parallel velocity. Properly accounting for the inhomogeneity of $\vec{B}_o$ yields a guiding center velocity $\vec{v}_{GC}$ that has a perpendicular contribution and results in the more general frequency  $\tilde{\omega}_N=\omega - \vec{k}.\vec{v}_{GC}-N\Omega$.

The generalization of the above expression to arbitrary harmonics is obtained by first adding the electric field inhomogeneity effect to the equation of motion which yields
\begin{equation} \label{Eq_mot}
x_{\perp_{1}}=x_{\perp_{1},o}+\frac{1}{2} \Big[  \frac{v_{+o}}{-i\Omega} e^{-i \Omega \tau} + \frac{v_{-o}}{i\Omega} e^{i \Omega \tau}  -\sum_N J_Nexp[-i\tilde{\omega}_N\tau] \Big ( \frac{\epsilon_+}{\tilde{\omega}_N(\tilde{\omega}_N-\Omega)}  + \frac{\epsilon_-}{\tilde{\omega}_N(\tilde{\omega}_N+\Omega)} \Big )   \Big ] $$
$$x_{\perp_{2}}=x_{\perp_{2},o}+\frac{1}{2i} \Big[  \frac{v_{+o}}{-i\Omega} e^{-i \Omega \tau} - \frac{v_{-o}}{i\Omega} e^{i \Omega \tau}  -\sum_N J_Nexp[-i\tilde{\omega}_N\tau] \Big ( \frac{\epsilon_+}{\tilde{\omega}_N(\tilde{\omega}_N-\Omega)}  - \frac{\epsilon_-}{\tilde{\omega}_N(\tilde{\omega}_N+\Omega)} \Big )  \Big ]  $$
$$x_{//}=x_{//,o}+v_{//o} \tau - \sum_N J_N exp[ -i \tilde{\omega}_N \tau] \frac{\epsilon_{//}}{\tilde{\omega}_N^2}
\end{equation}
in which the electric field is evaluated at the \textit{guiding center position} at $t=t_o$ and $v_{\pm,o}=v_\perp exp[\mp i \phi_o]$. Note that strictly
\textcolor{black}{$$v_{//o}=v_{//}(\tau=0)- \sum_N J_N \frac{i\epsilon_{//}}{\tilde{\omega}_N}.$$}
It is however customary to express the wave-induced changes w.r.t. the $unperturbed$ orbit i.e. to confuse $v_{//o}$ and $v_{//}(\tau=0)$. The obtained result is identical to the one earlier found provided $\omega \rightarrow \tilde{\omega}_N$ and adding the sum on the cyclotron harmonics and the associated Bessel functions. Higher order corrections when going from the particle to the guiding center position are omitted i.e.  $\partial \vec{x}/\partial \vec{x}_{GC}\equiv 1$ and thus $\nabla_{\vec{x}} \equiv \nabla_{\vec{x}_{GC}}$ although strictly both the Larmor radius and the unit vectors one projects on depend on the position.

Making the ansatz that driven quantities vary as a function of time as $exp[-i\omega t]$ renders real quantities complex. Products of such quantities are to be interpreted as products of their real parts: $AB \rightarrow Re[A]Re[B] = Re[A^*B+AB]/2$. This quantity has terms oscillating at twice the driver frequency and terms from which the time dependence has disappeared. At a frozen position, averaging over the fast time scale removes the term $Re[AB]/2$ and only retains the term $Re[A^*B]/2$. \textcolor{black}{The usual approach when averaging over 'fast time scales' is to average over the driver period as well as over the fast oscillatory aspects of the motion. As the bounce and toroidal drift motion are not included in the simple model adopted, only the average on the initial gyro phase $\phi_o$ needs to be done. Hence in this paper}
$$<...> = \frac{\omega}{2\pi} \int_0^{T=2\pi/\omega}d\tau \frac{1}{2\pi} \int_0^{2\pi} d\phi_o ...$$
for a given guiding center position. The next step is the integration over all possible parallel and perpendicular velocities for a prescribed distribution, assumed for simplicity to be a non-shifted Maxwellian $F_o=exp[-v^2/2v_t^2]/[(2\pi)^{3/2}v_t^3]$. The thermal spreading in the \textit{parallel} direction is crucial since it eliminates the resonant denominator that appeared in the cold plasma expression and replaces it by a smooth response. The following identities are useful when including the effects of the parallel dynamics:
$$\int_{-\infty}^{+\infty} dv_{//}exp[-\frac{v_{//}^2}{2v_t^2}]\frac{1}{v_{//}-v_{//m}}=\pi^{1/2}Z(\zeta_m)$$
$$\int_{-\infty}^{+\infty} dv_{//}exp[-\frac{v_{//}^2}{2v_t^2}]\frac{1}{(v_{//}-v_{//m})^2}=-\frac{(2\pi)^{1/2}}{v_t}[1+\zeta_mZ_m]$$
where $\zeta_m=v_{//m}/[2^{1/2}v_t]=[\omega-m\Omega]/[2^{1/2}k_{//}v_t]$. Finally the integration over the perpendicular velocities is performed. It requires using \cite{Watson,Gradshteyn}
$$\int_0^{+\infty} dv_\perp v_\perp exp[-\frac{v_\perp^2}{2v_t^2}] J_N^2=v_t^2 exp[-\lambda] I_N(\lambda),$$
where  $\lambda=\xi^2=(k_\perp v_t/\Omega)^2$. One finally gets
\begin{equation} \label{NEW_MAIN_EQ-2bis}
\vec{a}_{Pond}=\frac{1}{2} Re \Bigg [ \int d\vec{v} F_o <(\vec{x}-\vec{x}_{GC})^*.\nabla \vec{\epsilon}> \Bigg ] = $$
$$Re \Bigg [\sum_N \frac{exp[-\lambda]I_N(\lambda)}{2k_{//}v_t} \Bigg ( \frac{1}{\Omega 2^{3/2}}   \Big [ \Big ( \epsilon_{\perp_{1}}^*(Z_{N+1}-Z_{N-1})-i\epsilon_{\perp_{2}}^*(Z_{N+1}+Z_{N-1}-2Z_N) \Big )\frac{\partial \vec{\epsilon}}{\partial x_{\perp_{1}}} $$
$$ + \Big ( i \epsilon_{\perp_{1}}^*(Z_{N+1}+Z_{N-1}-2Z_N)+\epsilon_{\perp_{2}}^*(Z_{N+1}-Z_{N-1}) \Big ) \frac{\partial \vec{\epsilon}}{\partial x_{\perp_{2}}} \Big ]  + \frac{i\epsilon_{//}^* }{v_t} (1+\zeta_NZ_N) \vec{\epsilon} \Bigg ) \Bigg ].
\end{equation}
Together with the driver frequency, the cyclotron gyration frequency defines the fastest time scales in the problem. Averaging over all possible initial gyro-angles $\phi_o$ separates the various cyclotron harmonics $N$ such that a single and not a double sum on the cyclotron harmonics appears in the terms in which the electric field appears in front as well as behind the derivatives. As a result, the factor $exp[-i\tilde{\omega}\tau]$ is compensated by its complex conjugate so that all fast variation automatically disappears in these terms. For internal consistency $\Omega$ is considered constant here, as was assumed when integrating the equation of motion. When making this approximation, $\tilde{\omega}_N$ can simply be shifted $through$ the differential operators. Generalizing the found results requires accounting for the drift velocity caused by the magnetic field inhomogeneity and is not attempted here. The adopted philosophy in the present paper is that the dominant gradients are due to the RF electric field and not to gradients of equilibrium quantities. Such an assumption is clearly not always justified. In an $inhomogeneous$ plasma there is not a unique $\vec{k}$ and thus $\Psi$ and hence the direction in which the wave fronts advance cannot be neglected.

Note that the adopted approach only considers terms that are associated with the zero order distribution function; a supplementary set of terms (omitted here) arises from the density perturbation but does not require electric field inhomogeneity and thus does not yield a contribution in the here adopted approach which is based on a Taylor series development of the electric field. It goes without saying that further terms need to be added when background gradient effects are included as well. The expression discussed here is the generalization of the cold plasma expression and does $not$ require wave damping to realize a net flow.

\section{Adding the impact of the RF magnetic field}

So far, the RF magnetic field has been neglected. This is justifiable in the equation of motion of a charged particle undergoing the Lorentz force in the cold plasma limit and for fast waves in a not too energetic plasma. In order to describe the ICRH induced flows for arbitrary wavelengths and species energies, the magnetic corrections have to be added. Moreover, even if the magnetic field correction is small in the Lorentz force, this extra term is of similar magnitude as the already treated one when evaluating the $Ponderomotive$ force. Expressing the magnetic field in terms of the electric field via Maxwell's equation necessitates to add the term $1/2 Re[ <\vec{v}^* \times (\nabla \times \vec{\epsilon})/(i\omega)>]$ to the Ponderomotive force, which itself is a small correction. Similar algebra as what was done to find an expression for the electric field gradient contribution of the Ponderomotive acceleration, one obtains
\begin{equation}  \label{NEW_MAIN_EQ-3}
a_{Pond,\perp_{1}, B}=Re \textcolor{black}{\Bigg (}   \sum_N  \frac{exp[-\lambda] I_N(\lambda)}{2^{3/2}k_{//}v_t \omega} $$
$$\Bigg [ \frac{1}{2}\Bigg ( i \epsilon_{\perp_{1}}^*[Z_{N+1}-Z_{N-1}] + \epsilon_{\perp_{2}}^* [Z_{N+1}+Z_{N-1}] \Bigg )     \Bigg ( \frac{\partial \epsilon_{\perp_{2}}}{\partial x_{\perp,1}} - \frac{\partial \epsilon_{\perp_{1}}}{\partial x_{\perp,2}}\Bigg ) - \epsilon_{//}^* Z_N \Big ( ik_{//}\epsilon_{\perp,1}-\partial_{\perp,1}\epsilon_{//} \Big )  \Bigg ] \textcolor{black}{\Bigg )} $$
$$a_{Pond,\perp_{2}, B}= Re \textcolor{black}{\Bigg (} - \sum_N  \frac{exp[-\lambda] I_N(\lambda)}{2^{3/2}k_{//}v_t \omega} $$
$$\Bigg [ \frac{1}{2}\Bigg ( \epsilon_{\perp_{1}}^*[Z_{N+1}+Z_{N-1}] -i \epsilon_{\perp_{2}}^* [Z_{N+1}-Z_{N-1}] \Bigg )  \Bigg ( \frac{\partial \epsilon_{\perp_{2}}}{\partial x_{\perp_{1}}} - \frac{\partial \epsilon_{\perp,1}}{\partial x_{\perp_{2}}}\Bigg ) + \epsilon_{//}^* Z_N \Big ( ik_{//}\epsilon_{\perp_{2}}-\partial_{\perp,2}\epsilon_{//} \Big  )  \Bigg ] \textcolor{black}{ \Bigg )}$$
$$a_{Pond,//, B}= Re \textcolor{black}{\Bigg (}  \sum_N  \frac{exp[-\lambda] I_N(\lambda)}{2^{5/2}k_{//}v_t \omega} $$
$$\Bigg [ \Bigg ( \epsilon_{\perp_{1}}^*[Z_{N+1}+Z_{N-1}] -i \epsilon_{\perp_{2}}^* [Z_{N+1}-Z_{N-1}] \Bigg )  \Bigg ( ik_{//}\epsilon_{\perp_{1}}-\frac{\partial \epsilon_{//}}{\partial x_{\perp_{1}}}\Bigg )  + $$
$$ \Bigg ( i \epsilon_{\perp_{1}}^*[Z_{N+1}-Z_{N-1}] + \epsilon_{\perp_{2}}^* [Z_{N+1}+Z_{N-1}] \Bigg )
\Bigg (  ik_{//} \epsilon_{\perp_{2}}-\frac{\partial \epsilon_{//}}{\partial x_{\perp_{2}}} \Bigg )  \Bigg ] \textcolor{black}{\Bigg )}
\end{equation}
As before, the parallel component of the acceleration appears in the parallel equation of motion of the guiding center but will add contributions that are small compared to the thermal velocity, while the perpendicular acceleration - although small as well -  is more important since it will gives rise to a drift velocity across the magnetic field lines that is absent when there is no ICRH electric field.

\section{A side note on omitted rapidly varying linear terms}

Ponderomotive acceleration in a tokamak is the net residual acceleration caused by an inhomogenous, rapidly varying electric field $along$ the orbit after all fast time scale effects due to both the driver and the confining magnetic field have been removed. The fact that the expression for the electric field at the guiding center position still contains the time factor $exp[-i\omega t_o]$ (where $t_o$ is interpreted as the reference time for the fast and $the$ time on the slow time scale) causes terms in the Ponderomotive force that are linear in $\epsilon_\alpha$ rather than quadratic and of the form $\epsilon_\alpha \epsilon_\beta^*$ to $still$ be oscillating fast as a function of $t_o$. Such terms appear as a consequence of the fact that the displacement nor the velocity is simply proportional to the electric field but has supplementary terms due to the presence of the strong magnetic field. In the classical treatment of the Ponderomotive force, formulated in absence of a strong magnetic field and thus formulated for the case where there was only 1 relevant fast scale,  such terms do not appear. Rather than giving rise to a set of contributions at the various cyclotron harmonics, they isolate 1 harmonic in particular per term. As these contributions do not represent 'slow time scale' dynamics, a supplementary average thus needs to be performed to remove this remaining fast behavior when advancing along the orbit. Because the start and end point for the $\tau$ integral appearing in the previous 2 sections can be chosen in a number of ways ($t_o \le t \le t_o+T$, $t_o -T/2 \le t \le t_o+T/2$, ...), treating all possible choices as equally probable is equivalent to performing the supplementary $t_o$ average aside from the $\tau$ average. This extra average is clearly unnecessary for the quadratic terms with complex conjugate phase factors since $t_o$ dependence is already absent in these. For the remaining terms integrating over a driver period not only removes the oscillatory behavior but removes the terms altogether. As a result the $usual$ substitution $AB \rightarrow Re[A^*B]/2$ could be adopted before. This is not equivalent to stating that these deviations from the average do not contain important physics. It merely states that these oscillatory terms are irrelevant when studying slow time scale dynamics. In this section we spend a few words on what these contributions look like.

Because of the assumption that the static magnetic field is constant both in amplitude and direction so that the guiding center motion only has a parallel component, the relevant type of integral prior to the $t_o$ averaging is of the form
$$ I_{//}(\xi,p_o)= \int_{-\infty}^{+\infty}dv_{//}exp[-\frac{v_{//}^2}{2v_t^2}] \frac{1}{T} \int_0^T d\tau exp[-i \tilde{\omega}_m\tau]= $$
$$\int_{-\infty}^{+\infty}dv_{//}exp[-\frac{v_{//}^2}{2v_t^2}] \frac{\omega}{-i \tilde{\omega}_m2\pi}[exp[-i \tilde{\omega}_m\frac{2\pi}{\omega}]-1]$$
where $T=2\pi/\omega$ is the period associated with the driver frequency. When $\tilde{\omega}_m=M\omega$ where $M$ is an integer, then the integral vanishes. When $\tilde{\omega}_m/\omega$ is real but non-integer, the result is finite. The only possibility to make the period average constant and finite is to choose $\tilde{\omega}_m=0$. The limit $\tilde{\omega}_m\rightarrow 0$ of the time integral is $1$ and isolates a $v_{//}$ where the velocity integrand picks up its major contribution. Since $\tilde{\omega}_m$ is of the form $-k_{//}(v_{//}-v_{//,res})$ where $\tilde{\omega}_m=0 $ for $v_{//}=v_{//,res}$ allows to find
 \begin{equation} \label{Intpar}
 I_{//}(\xi,\zeta)= \frac{1}{i\xi} \Big [  \pi^{1/2}(exp[\eta^2-\zeta^2]Z(\eta)-Z(\zeta))  + P \Big ]
 \end{equation}
 where $\xi=2\pi k_{//}/\omega$, $\zeta=v_{//,res}/[2^{1/2}v_t]$ and $\eta=\zeta-i\tilde{\xi}/2$ with $\tilde{\xi}=2^{1/2}v_t \xi$
 while $Z$ is the plasma dispersion function. Note that $\eta$ is complex and requires evaluating the Fried-Conte functions for non-real argument. After the coordinate transformation the contour has been deformed at infinity to recast it on the real axis. Adopting the usual causality rule $\omega \rightarrow \omega + i \nu$ where $\nu$ is infinitesimally small and positive (physically representing a collision frequency), this gives rise to the pole extra pole contribution $P$: $P=2 \pi i exp[-\eta^2]$ when $k_{//} > 0$ and $P=0$ when $k_{//} < 0$. Increasing $\xi$  gradually isolates the resonance location in $I_{//}$ as being the only location where the integral picks up a significant net contribution i.e. approaching the behavior of a delta function for the oscillatory factor. The integral shrinks for increasing $\xi$.

It can readily be seen that the relative importance of these oscillating terms and the ones that survive the average over $t_o$ depends on the relative magnitude of the electric field and of the thermal velocity. For example the $<(\vec{x}-\vec{x}_{GC})^*.\nabla \vec{\epsilon}>$ term then is supplemented by the extra term

\begin{equation} \label{NEW_MAIN_EQ-2bissss}
\frac{1}{2} Re \Bigg [ \int d\vec{v} F_o <(\vec{x}-\vec{x}_{GC})^*.\nabla \vec{\epsilon}> \Bigg ]_{extra} = $$
$$Re \Bigg [ \frac{ik_\perp v_t}{2^{3/2}\pi^{1/2}\Omega^2}I_{//}(\xi,\zeta_o) exp[-\lambda/2] (cos\Psi \frac{\partial}{\partial x_{\perp,1}} + sin\Psi \frac{\partial}{\partial x_{\perp,2}})\vec{\epsilon} \Bigg ]
\end{equation}

\section{Examples}

As an example, the ICRH induced drift velocities consistent with a ICRH $(^3He)-H$ fundamental cyclotron frequency heating scenario typical for JET are provided. A concentration $N_{^3He}/N_H=2\%$ was assumed. A scan over various important parameters is presented. The present computations where performed for a given electric field magnitude of $4.6kV/m$. To the exception of the magnetic field scan (first 4 figures), the polarization is assumed to be fixed. For the perpendicular components the choice $E_{\perp,1}=3kV/m$,  $E_{\perp,2}=i 3.5kV/m$ was then made. For the parallel electric field $E_{//}=20V/m<<E_\perp$, a value both typical for the fast wave and for electrostatic modes, was chosen. Unless specified otherwise, the parallel wave number is taken to be $k_{//}=6/m$, the magnetic field strength is $B_o=3.25T$, the driver frequency is $f=33MHz$ and the wave fronts are aligned with the first perpendicular direction, $\Psi=0$, while the temperature is $T=3keV$. Cyclotron harmonics up to $|N|=4$ were retained. For the adopted parameters, the fast wave carries $0.35MW$ when launched from an antenna with a surface of $1m^2$.

\begin{figure} [lt]
\centering
\includegraphics[width=5in]{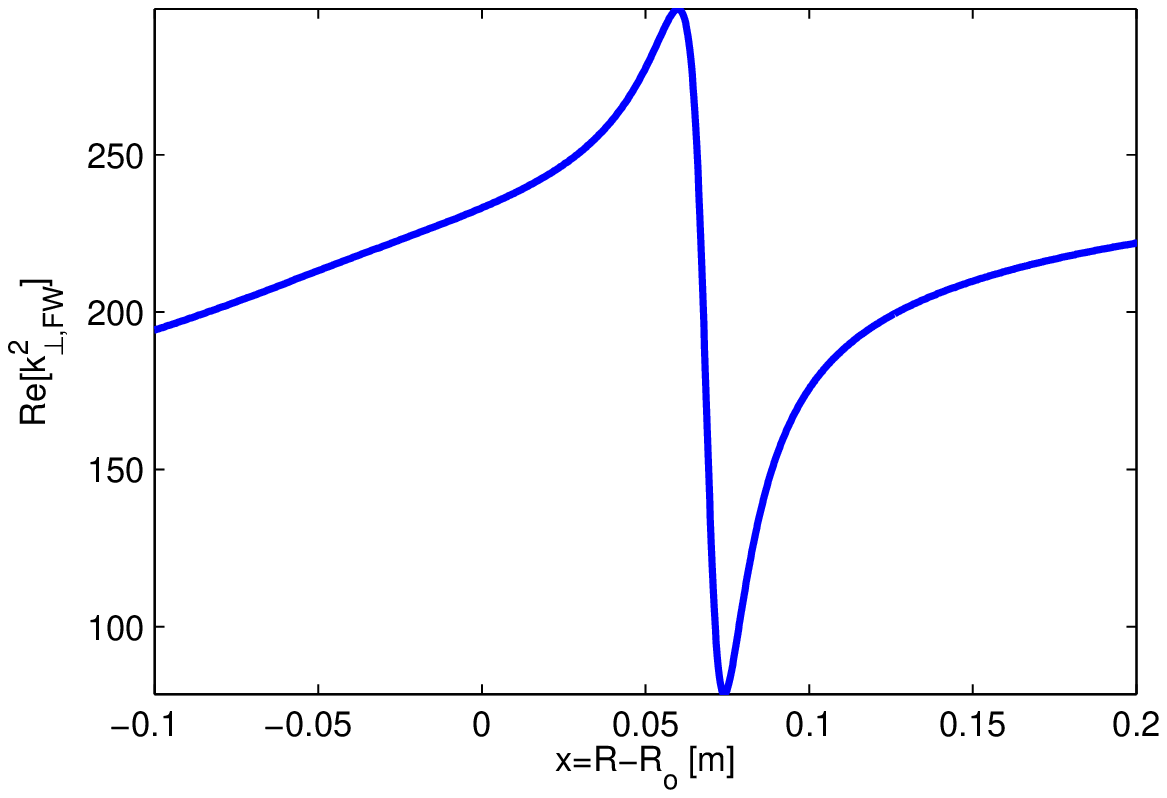}

\includegraphics[width=5in]{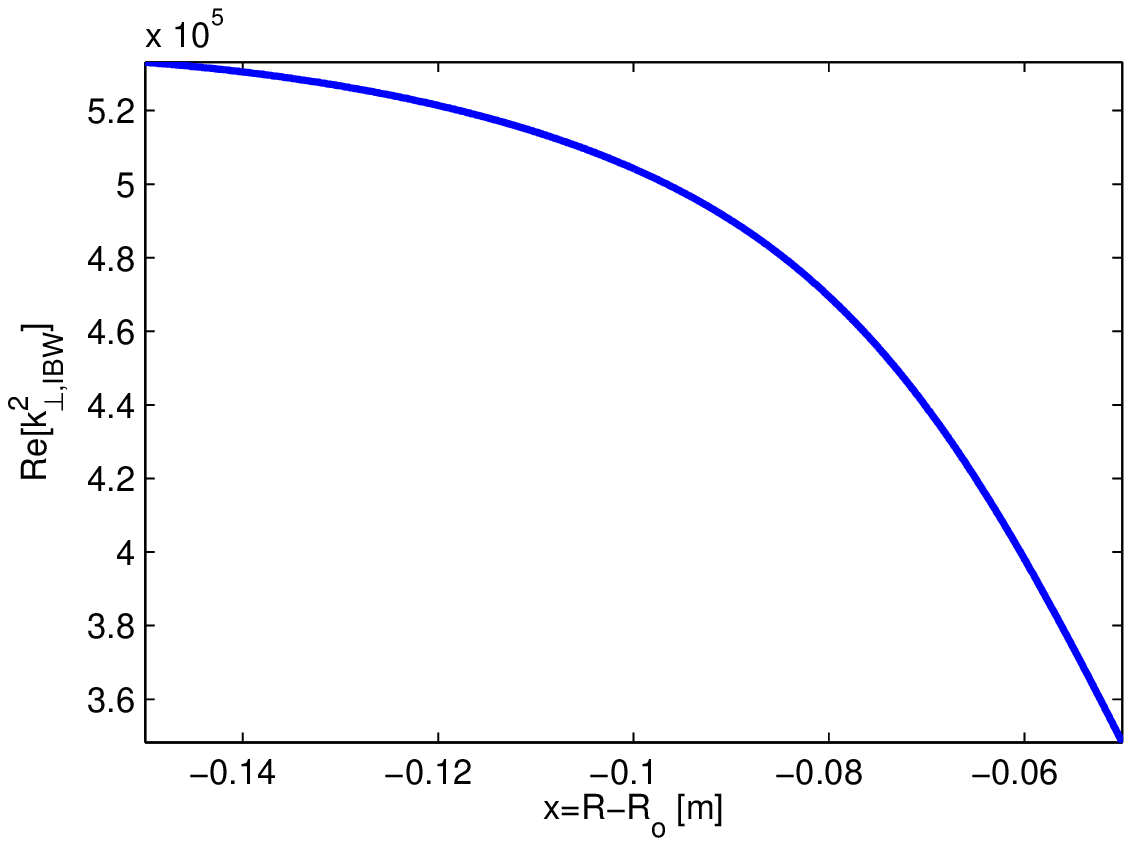}
\caption{Real part of $k_\perp^2$ for the fast (top) and the Bernstein (bottom) wave dispersion equation root in a $(^3He)-H$ plasma with $N_{^3He}/N_H=2\%$, and a parabolic temperature as well as density profile with central temperature of $T_o=3keV$ and density of $N_o=3\times 10^{19}/m^3$; $B_o=3.25T$, $f=33MHz$, $k_{//}=6/m$ and $\Psi=0$. The Bernstein wave root has only been plotted in the (high field side) region where it is propagative. }
\label{FIG_kperp2}
\end{figure}

\begin{figure} [lt]
\includegraphics[clip,height=8cm,trim=2.7cm 0cm 0 0]{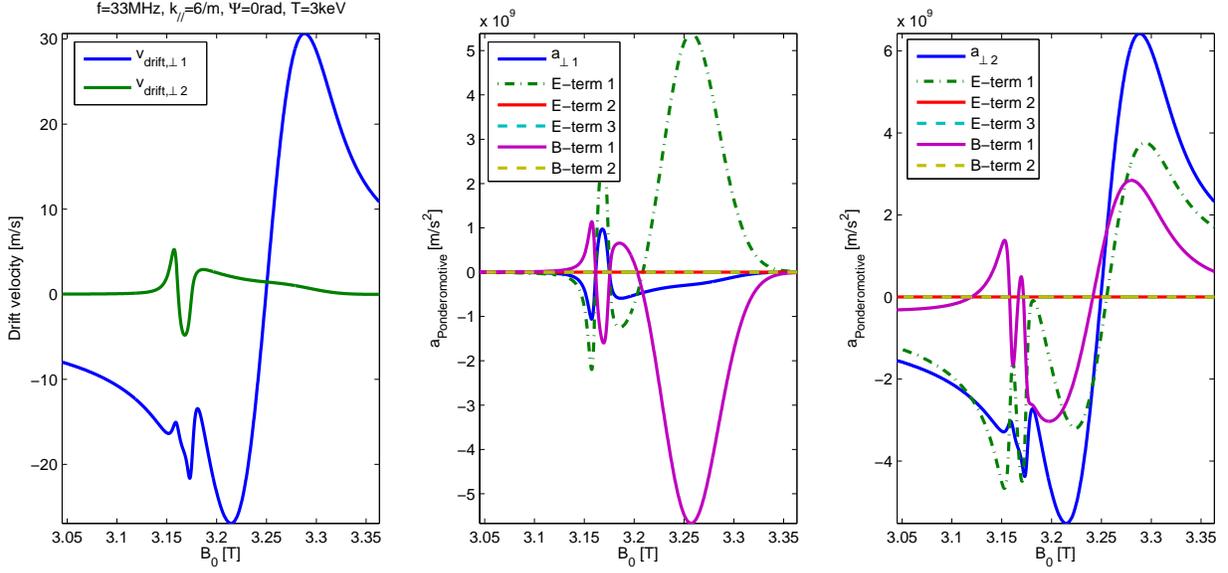}
\caption{Scan of the magnetic field strength for fixed frequency $33MHz$, parallel wave number $6/m$ and wave front angle $\Psi=0$ for the fast wave dispersion equation root depicted in the Fig. \ref{FIG_kperp2} and the corresponding polarization in a $(^3He)-H$ plasma with parameters as given before. The magnitude of the electric field strength is $|E|=4kV/m$. As $B_o$ grows as $1/R=1/[R_o+x]$, the left side of Fig. \ref{FIG_kperp2} corresponds to the right side of the present figure.}
\label{FIG_ikind1FW}
\end{figure}

\begin{figure} [lt]
\includegraphics[clip,height=8cm,trim=2.7cm 0cm 0 0]{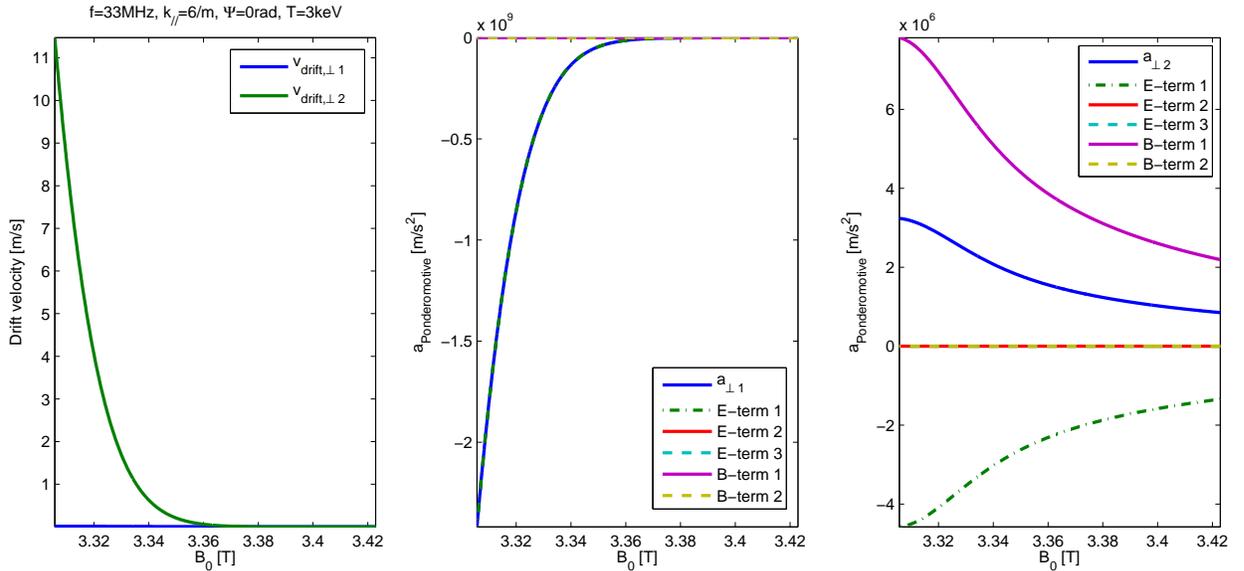}
\caption{Scan of the magnetic field strength for fixed frequency, parallel wave number and wave front angle for the ion Bernstein wave dispersion equation root depicted in the Fig. \ref{FIG_kperp2} and the corresponding polarization in a $(^3He)-H$ plasma with parameters as given before. As $B_o$ grows as $1/R=1/[R_o+x]$, the left side of Fig. \ref{FIG_kperp2} corresponds to the right side of the present figure.}
\label{FIG_ikind1IBW}
\end{figure}

Figure \ref{FIG_kperp2} depicts the fast and Bernstein wave roots computed with an all order finite Larmor radius dispersion equation solver based on the dielectric tensor as given in Stix \cite{Stix}. The top subfigure is the fast wave root in a region including the mode conversion point to the Bernstein wave. The bottom subfigure is the Bernstein wave root in the region on the high field side of the confluence i.e. the region where the Bernstein wave is propagative. Figure \ref{FIG_ikind1FW} depicts how the drift velocities and the amplitude of the Ponderomotive acceleration change as a function of the magnetic field strength for the fast wave, accounting for the proper polarization of the wave computed with the dispersion equation root depicted in the previous figure; remind that the magnetic field increases for decreasing major radius so the left side of the plot corresponds to the right side of the dispersion plot. The left subfigure depicts the drift velocities. As the direction of the phase velocity and that of the group velocity are almost parallel for the fast wave, the first perpendicular direction can be interpreted as the radial direction and the second as the poloidal direction for a typical fast wave antenna with its straps aligned with the magnetic surfaces launching waves nearly radially into the plasma. The dominant velocity component is in the radial direction. Although the drift velocities are small with respect to the thermal velocity, they cannot be neglected since the guiding center motion in absence of the waves does not even have a component that moves particles across the magnetic surfaces. Hence the drift velocities shown are to be compared with pinch velocities arising from transport, the latter typically being of the order of a few $m/s$. For typical electric field strengths, the wave induced drifts dominate the transport drifts and cannot be neglected. The middle and right subfigures show the perpendicular components of the Ponderomotive acceleration and the terms they are composed of; the numbering corresponds to the order in which the terms appear in equations \ref{NEW_MAIN_EQ-2bis} and \ref{NEW_MAIN_EQ-3}. As mentioned when deriving the equations, due to the presence of the strong magnetic field  the $poloidal$ component of the acceleration gives rise to a $radial$ drift. Note that the radial component of the acceleration has a contribution from the electric field and from the magnetic field that almost balance one another, while they tend to re-enforce one another for the poloidal component. Figure \ref{FIG_ikind1IBW} depicts how the drift velocities and the amplitude of the Ponderomotive acceleration change as a function of the magnetic field strength for the Bernstein wave. Due to the fact that the Bernstein wave essentially is an electrostatic mode, its electric field is aligned with its wave vector and thus $E_{\perp_1}$ is much larger than $E_{\perp_2}$. As a consequence the wave induced drift is essentially in the direction perpendicular to the direction of the wave front propagation. Close to the ion-ion hybrid layer at which a radially propagating fast wave excites the Bernstein wave, the resulting drift is in the poloidal direction. However, the direction in which the Bernstein wave moves sensitively depends on the poloidal magnetic field and tends to turn away from the direction in which it was initially propagating; away from the equatorial plane where poloidal field effects dominate temperature effects, the Bernstein wave becomes an ion cyclotron wave. At that stage the induced drifts are expected to have a poloidal as well as a radial component.

\begin{figure} [lt]
\hspace{-50pt}
\includegraphics[width=4.in]{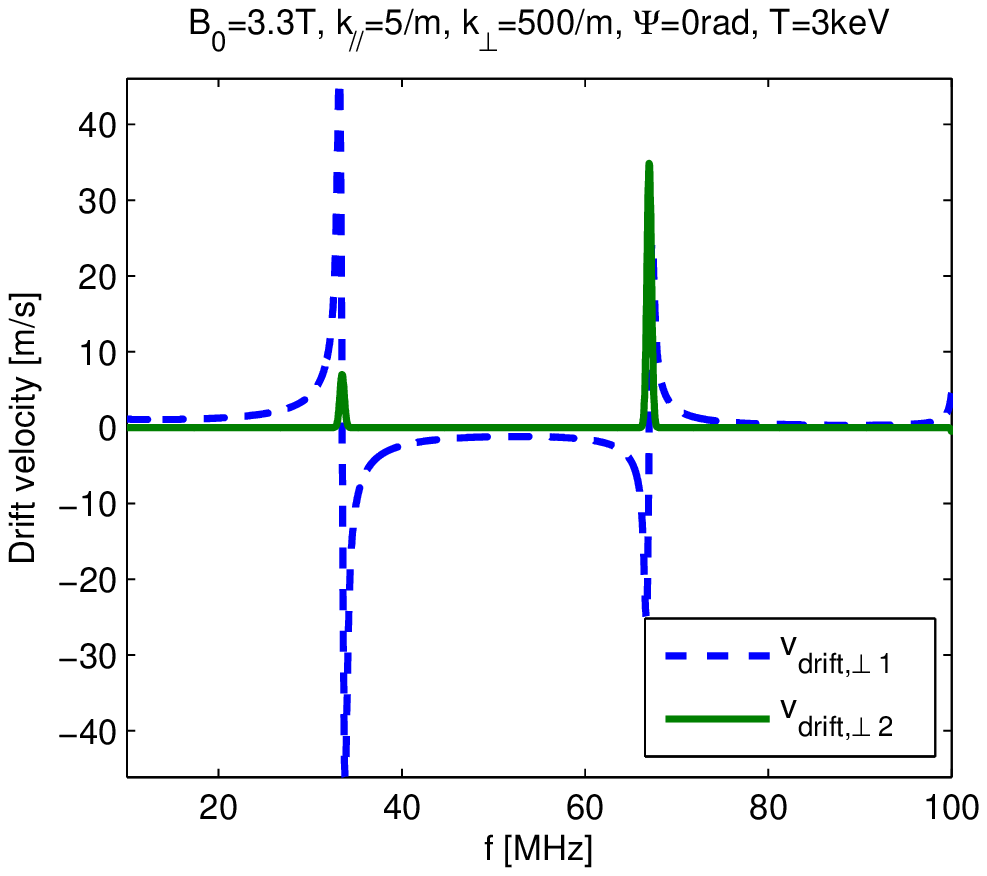}
\includegraphics[width=4.in]{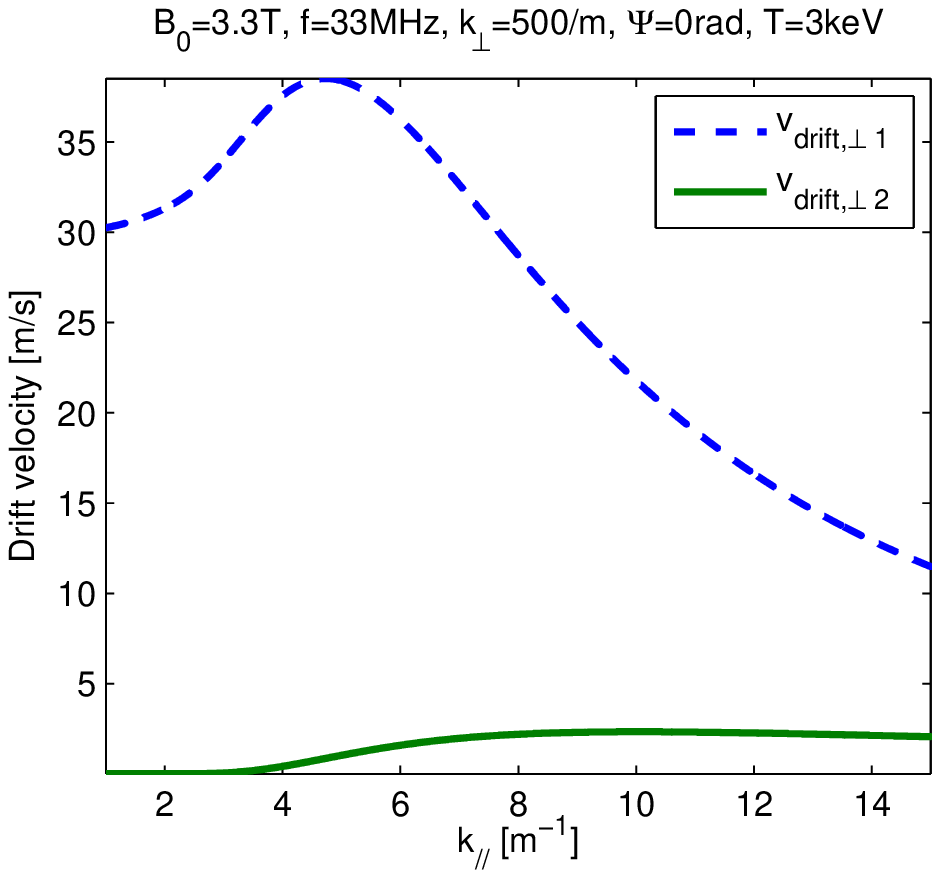}

\hspace{-50pt}
\includegraphics[width=4.in]{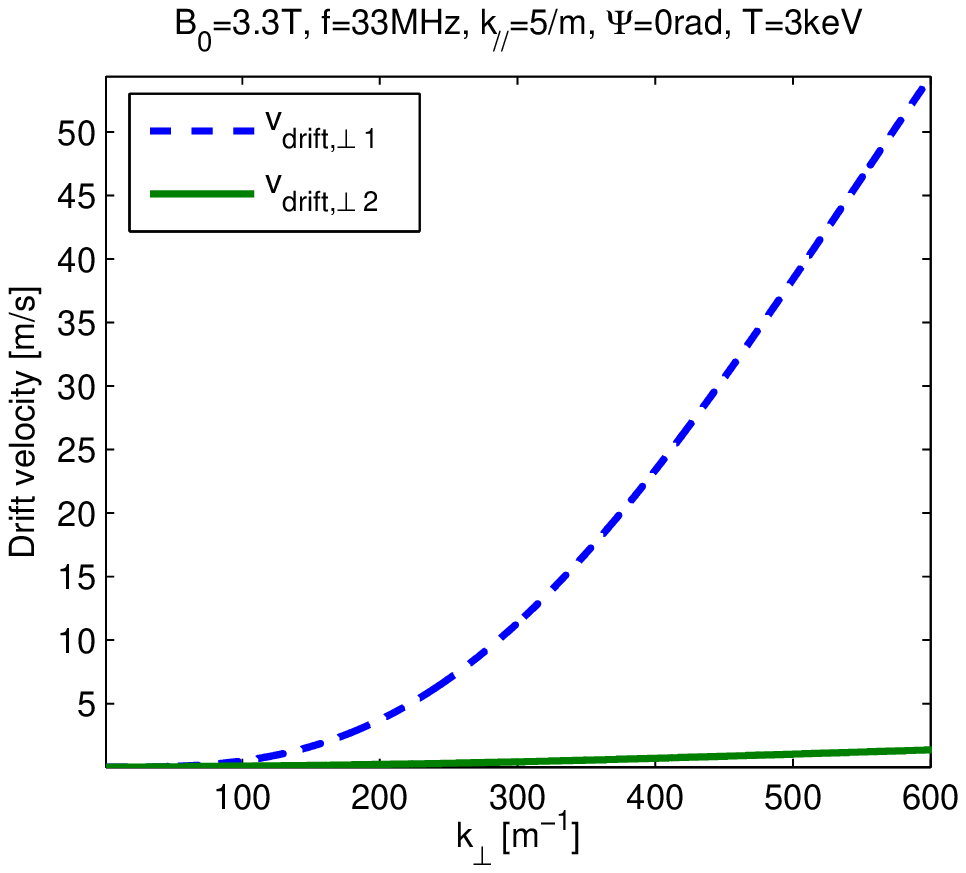}
\includegraphics[width=4.in]{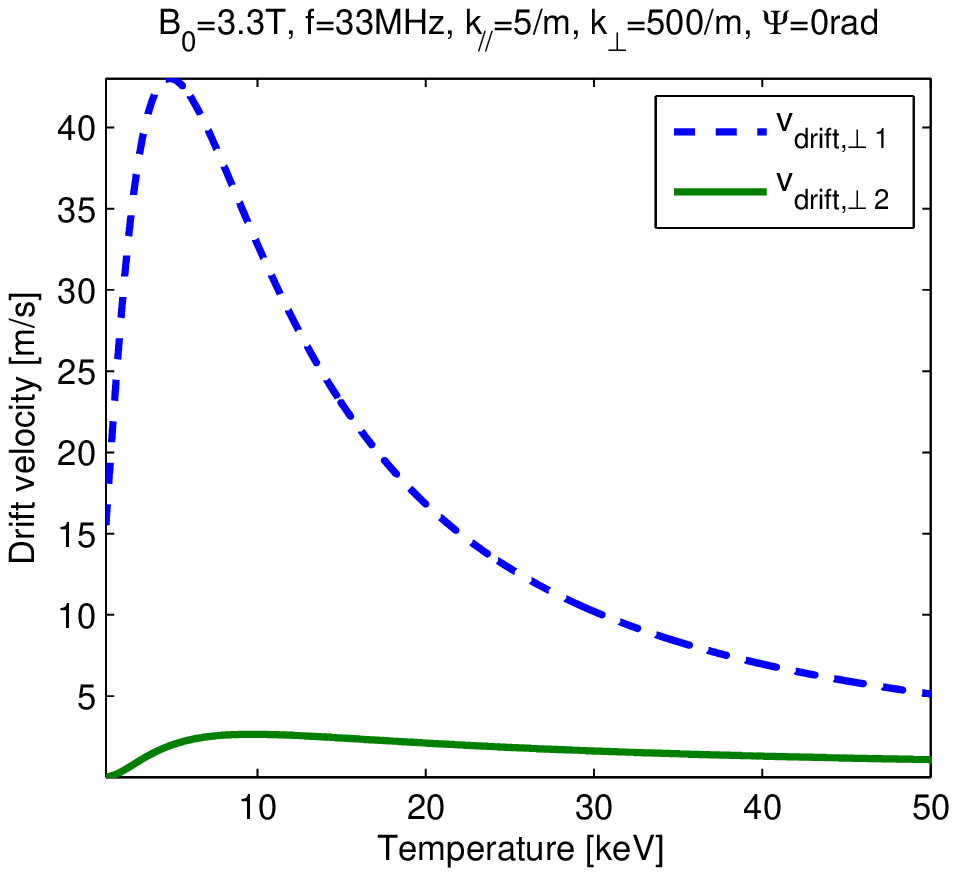}
\caption{Perpendicular drift velocity as a function of the driver frequency, the parallel wave number, the perpendicular wave number and the temperature using a fixed electric field strength; $B_0=3.3T$ and $k_{//}=5/m$ except in the $k_{//}$ scan.}
\label{FIG_ikind2}
\end{figure}

Figure \ref{FIG_ikind2} shows the dependence of the perpendicular drift velocities  on the frequency, parallel wave number, perpendicular wave number and temperature for fixed magnetic field strength, density and wave front direction. In the top left subplot, it can be seen that at opposite sides of cyclotron layers, the drift velocity component aligned with the wave fronts has opposite sign, while the other component is only of importance very close to the resonances. The right top subplot shows the dependence of the perpendicular drift velocities  on the parallel wave number. Except at small values, the amplitude of the perpendicular drift velocity decreases when $k_{//}$ grows. The opposite is the case as a function of the perpendicular wave number. The bottom left subplot shows the dependence of the perpendicular drift velocities on this parameter. From this scan one tends to conclude that the impact of short wavelength branches is much more significant than that of long wavelength modes. However, this statement needs to be moderated since the values depicted are obtained assuming the electric field is $given$ while in practice the flux carried by the wave is the more relevant parameter when tracing the fate of power carried by waves. Finally, the bottom right plot shows the dependence of the perpendicular drift velocities  on the temperature. The higher the temperature, the lower the induced drifts.

\section{Discussion}

At the outset, the equation of motion solved in the present paper assumes particles execute their Larmor gyration around a guiding center that is glued to field lines of a constant confining magnetic field. It is intended to provide the dominant effect of drifts generated by ICRH waves but e.g. omits the impact of deviations of particles to regions with significantly different densities and/or temperatures. Strictly, the above found expression allows to assess the importance of ICRH induced flows at arbitrary harmonics but $only$ for a homogeneous plasma and for an electric plane wave field with a single mode $\vec{k}$. It allows to get an idea of the magnitude of the induced drift velocity for a given electric field amplitude. Averaging over the full distribution resulted in the removal of the ambiguity of what happens at the cyclotron resonance and replaced the cold plasma divergence of the earlier obtained flow velocity by a smooth transition where flow is efficiently driven close to the cyclotron layers. When the drift velocity obtained by evaluating the derived expressions is excessive, there is no justification for the use of the adopted equation of motion and a more rigorous development is required.

By neglecting inhomogeneity corrections, the computations have been reduced to a bare minimum in this paper. The first of such a variation that comes to mind is that of the confining magnetic field in presence of a plasma current. Rather than straight lines, the field lines then are helicoidally twisted lines on magnetic surfaces. Clearly in that case - e.g. because $v_{//}$ is no longer a constant - it makes little sense to claim the present model can predict the average flow on a magnetic surface. Nevertheless, one may hope the order of magnitude the model predicts has sufficient physical meaning to provide an indication on whether ICRH induced drifts should be incorporated in rigorous modeling or not. Since the density and the temperature are constant on magnetic surfaces, it seems natural to define the unit vector $\vec{e}_{\perp_{1}}$ as perpendicular to magnetic surfaces, and $\vec{e}_{\perp_{2}}$ - up to a small correction because of the finite poloidal field - as pointing in the poloidal direction. To leading order, the guiding center then does not feel density or temperature inhomogeneity effects. Because the toroidal magnetic field dominates the poloidal field, and since the former varies as $1/R$, with R the major radius, a second plausible option is to identify $\vec{e}_{\perp_{1}}=\vec{e}_R$ and $\vec{e}_{\perp_{2}}=\vec{e}_{//}\times \vec{e}_R\approx \vec{e}_Z$. For waves launched in large machines using antennas that are close to the equatorial plane, the 2 choices are equivalent as the magnetic field gradient is aligned with that of the density and temperature; $v_{drift,\perp_{1}}$ then is a radial and $v_{drift,\perp_{2}}$ a poloidal drift. When the wave fronts are oriented differently, this identification is no longer justified.

Both through the dependence of $\lambda$ on the background variation as through the spatial inhomogeneity of the RF electric field, it can be seen that short wavelength waves seem more likely candidates to drive flow than long wavelength branches. As the flow velocities are fairly small w.r.t. the thermal velocities, Coulomb collisional cross-talk between various types of species is a higher order effect that can be neglected i.e. the obtained flows do not (strongly) invalidate the implicit assumption that the distribution functions are Maxwellians. On the other hand, the thermal motion will tend to spread out the effect of the flow over orbits. In absence of drift orbit corrections - an assumption implicitly made when reducing the drift velocity to the parallel component $v_{//}$ - this spreading, and the guiding center poloidal motion itself tend to average the flow over magnetic surfaces but - although small - radial corrections remain. The up-down asymmetry of the wave behavior on account of the finite poloidal field forces one to rigorously account for the short wavelength wave pattern. Provided the magnitude of the resulting flow is indeed small compared to the thermal velocity, the thus obtained average flow can - to leading order - be neglected when solving the wave equation, but the flow has to be accounted for when describing the slower time scale dynamics. Furthermore, should the flow velocity not be small enough (there is no indication of that being the case in the examples treated), it is necessary to solve the wave equation properly accounting for the shifted distribution which significantly increases the complexity of the problem. To date, only a very limited amount of wave solvers are able to rigorously account for arbitrary distributions (see e.g. \cite{Jaeger}), and none of them are convincingly converged for the short wavelength branches that are thought to be crucial for ICRH induced flows.

In spite of the found flow velocities being relatively small compared to the thermal velocities, they are $not$ at all small compared to the convection speeds appearing in transport models. In the same limit as what was assumed in the present paper when solving the equation of motion (guiding centers simply following magnetic field lines since gradients of the static magnetic field have been dropped w.r.t. the stronger gradients arising from the electric field inhomogeneity) the poloidal flows that are induced are believe to merely represent a $correction$ to the already existing motion of the population. However, terms that are $bound$ to have an impact on a more readily noticeable scale are the radial flows, which exceed the speed at which populations move across magnetic surfaces due to transport effects. These radial drifts are a natural candidate for explaining wave heating induced confinement degradation, but it is premature to make a firm statement on this matter as the here described drifts cause a different density migration of different species (as can easily be seen by introducing the obtained drifts in the continuity equation of the species and was already shown - be it in the context of ICRH induced density depletion close to antennas - in the cold plasma limit discussed in \cite{DVE_Ne_depletion}) and thus create charge imbalances which the plasma will try to restore by creating a potential. Although crucial, this detailed slow time scale dynamics is beyond the scope of the present paper.

The term 'Ponderomotive force' is not always interpreted in the same way, although the basic philosophy is common: it is the slow time scale force that results from averaging over all fast variations when feeding the results of a linear approximation back into the equation of motion to obtain a second order correction that can be added to the equation of motion describing the slow time scale dynamics. A different set of terms altogether from the ones discussed in the present paper results when including the \textit{perturbed density} into the formulation. As was shown by Myra and D'Ippolito \cite{MyraDIppolito}, the time derivative of the perturbed density ($\partial N_1/\partial t \rightarrow -i \omega N_1$) and accounting for the second order terms arising from the perturbed density and the perturbed distribution function $f_1$ results in Ponderomotive force terms that require \textit{damping} (the $imaginary$ part of the Fried-Conte function) to be present in order for flows to be created, an aspect that is at odds with the momentum transfer already exhibited in the cold plasma version of the Ponderomotive force and hence basically different from the effect discussed here. The acceleration identified in the present paper differs from but is of similar amplitude as that identified by Myra et al. On the other hand, both the direction and the magnitude of the associated drifts differs because the equation of motion is solved in presence of the strong magnetic field and thus yields terms that dominate the 'friction' terms invoked in the work of Myra to reach a steady state solution. The 2 types of terms (resonant interaction requiring damping and non-resonant interaction which does not) are discussed by Hellsten \cite{Hellsten_rotation}, making a number of approximations to keep the discussion intuitive. His work shows the subtlety of the description and the need for an ordering. Although strictly speaking not consistent with the adopted solution of the equation of motion, Hellsten includes cyclotron frequency $variation$ terms in the expression of the Ponderomotive force that undoubtedly play a role in the wave-particle interaction. In Hellsten's work the implicit assumption is thus made that the scale length of the fields and that of the background are both important and thus need to be retained. Omitting this effect in the equation of motion likely implicitly relies on the fact that the omitted drifts are in an ignorable perpendicular direction for the slab results presented and that the wave vector component in the ignorable perpendicular direction is small so that it e.g. does not significantly modify the resonance condition.

Of course, $all$ linear variations that can give rise to higher order corrections that contain a slowly varying component should be included and thus the here provided expressions are only a subset of the net corrections on the slow time scale of the fast time scale dynamics.

\section{Conclusion}

A model was presented to assess the flows induced by ICRH waves in the presence of a strong magnetic field. These flows are the finite temperature counterpart of flows already existing in cold plasmas and do not require the waves to be damped. The induced flows are perpendicular both to the Ponderomotive acceleration and to the magnetic field lines. Hence they modify the equilibrium. Via a few examples, it was shown that for realistic field magnitudes, the created flows are small compared to the thermal velocities (allowing to neglect the impact of the flow on the distribution function and thus on the dielectric response of the plasma to the wave) but are large compared to the intrinsic flow rates connected to transport phenomena. In particular, the described flows are believed to add mere corrections to the poloidal flows but introduce radial flows that are expected to influence the plasma equilibrium. The proper treatment of the latter is outside the scope of the present paper but needs to be addressed.

\newpage

\end{document}